\documentclass[a4paper]{jpconf}
\usepackage{graphicx}
\begin{document}
\title{Production of Strange, Non-strange particles and Hypernuclei in an Excluded-Volume Model}

\author{S. K. Tiwari, and C. P. Singh}

\address{Department of Physics, Banaras Hindu University, Varanasi-221005, India}

\ead{sktiwari4bhu@gmail.com}

\begin{abstract}
We present a systematic study of production of strange and non-strange hadron yields and their ratios obtained in various experiments using our thermodynamically consistent excluded-volume model. We also analyze the production of light nuclei, hypernuclei and their antinuclei in terms of our excluded-volume model over a broad energy range starting from Alternating Gradient Synchrotron (AGS) to Large Hadron Collider (LHC) energies. Further, we extend our model for studying rapidity spectra of hadrons produced in heavy-ion collisions.  
\end{abstract}

\section{Introduction}
The production of strange particles in heavy-ion collisions is considered as one of the most important signals for the phase transition from hot, dense hadron gas (HG) to quark-gluon plasma (QGP) \cite{singh:1993}. Recently we have analyzed the production of strange hadrons over a broad energy range from AGS to LHC energies using our new thermodynamically consistent excluded-volume model \cite{Tiwari:2012}. We have suitably incorporated the attractive interactions between hadrons through the addition of hadron resonances in our model. We assign an equal hard-core radius to each baryon in order to include repulsive interactions between them while mesons in the model are treated as pointlike particles. Using chemical freeze-out criteria as proposed by us \cite{Tiwari:2012}, we calculate strange and non-strange hadron ratios and compare our results with the experimental data. We also study the production of light nuclei, hypernuclei and their antinuclei from AGS to LHC energies using our model and confront our model predictions with the experimental results. Moreover, we analyze the experimental data on rapidity distributions of various hadrons in terms of our model and elucidate the role of the longitudinal flow present in the medium formed in heavy-ion collisions. In the last, we present the conclusions and summary.

\section{Formulation of Model}
The grand canonical partition function for baryons in our thermodynamically consistent excluded-volume model can be written as follows \cite{Tiwari:2012,singh:2009} :   

\begin{equation}
ln Z_i^{ex} = \frac{g_i}{6 \pi^2 T}\int_{V_i^0}^{V-\sum_{j} N_j V_j^0} dV
\int_0^\infty \frac{k^4 dk}{\sqrt{k^2+m_i^2}} \frac1{[exp\left(\frac{E_i - \mu_i}{T}\right)+1]},
\end{equation}
where $g_i$ is the degeneracy factor of ith species of baryons, $E_{i}$ is the energy of the ith particle ($E_{i}=\sqrt{k^2+m_i^2}$), $V_i^0$ is the eigen volume assigned to each baryon of ith species. We have assigned an equal volume $\displaystyle V^{0}=4\pi r^{3}/3$ for each baryon with a hard-core radius $r=0.8\; fm$. Mesons in our model are considered as pointlike as they can penetrate each others volume. We have included all the baryons, mesons and their resonances having masses upto $2\;GeV/c^{2}$ in our model. To ensure strangeness conservation, we have imposed the condition of strangeness neutrality by considering $\sum_{i}S_{i}(n_{i}^{s}-\bar{n}_{i}^{s})=0$ where $S_{i}$ is the strangeness of ith hadron.

Recently, Gorenstein \cite{Gorenstein:2012} has criticized our model on the basis of thermodynamical and statistical consistency and claimed that the thermodynamical average of number of baryons is not equal to its statistical average in our model. We calculate these two averages in our model and find a relationship between them as follows \cite{Tiwari:2013} :
\begin{equation}
\overline{n}= \langle n \rangle \; - T \langle \frac{n^{0}}{(1-R)} V^0 \frac{\partial \overline{n}}{\partial \mu}~\rangle.
\end{equation}
We see that the thermal average of number of baryons is not exactly equal to its statistical average in our model and an extra term appears because $\overline {N}$, present in the available volume ($V~-~V^0\overline{N}$), is a function of $\mu$. We call this extra term as ``correction term'' and plotted this term and $\overline {n}$ with respect to $T$ at constant $\mu_B=400 MeV$ in Fig. 1. We find that ``correction term'' gives an almost negligible contribution and hence can be ignored. Thus, our model can safely be considered as thermodynamically consistent \cite{Tiwari:2013}.

\begin{figure}[h]
\includegraphics[width=14pc]{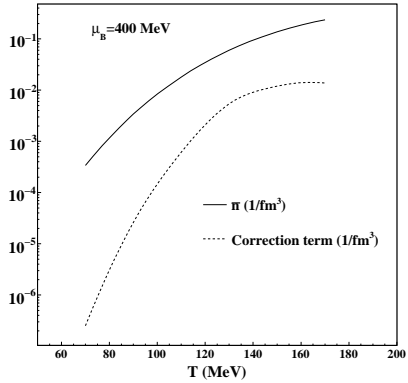}\hspace{2pc}%
\begin{minipage}[b]{14pc}\caption{\label{label}Variation of thermodynamical average of the number density of baryons and the ``correction term'' with respect to $T$ at constant $\mu_B=400\;MeV$ \cite{Tiwari:2013}.}
\end{minipage}
\end{figure}


 The rapidity distributions of baryons can be calculated using our thermal model as follows \cite{Tiwari1:2013} :

\begin{equation}
\Big(\frac{dN}{dy}\Big)_{th}=\frac{g_iV\lambda_i}{2\pi^2}\;\Big[(1-R)-\lambda_i\frac{\partial{R}}{\partial{\lambda_i}}\Big]
exp\left(\frac{-m_i\;coshy}{T}\right)\Big[m_i^2T+\frac{2m_iT^2}{coshy}+\frac{2T^3}{cosh^2y}\Big].
\end{equation}

where $m_i$ is the mass of the ith species and $V$ is the freeze-out volume of the system. After incorporating the longitudinal flow in our thermal model the formula for the rapidity distributions of baryons takes the following form \cite{Tiwari1:2013} :

\begin{eqnarray}
\frac{dN_i}{dy}=\int_{-\eta_{max}}^{\eta_{max}} \Big(\frac{dN_i}{dy}\Big)_{th}(y-\eta)\;d\eta,
\end{eqnarray}
where $\eta_{max}$ is a free parameter related with the longitudinal flow velocity ($\beta_L$) \cite{Tiwari1:2013}.

\section{Results and Discussions}

\begin{figure}[h]
\begin{minipage}{14pc}
\includegraphics[width=14pc]{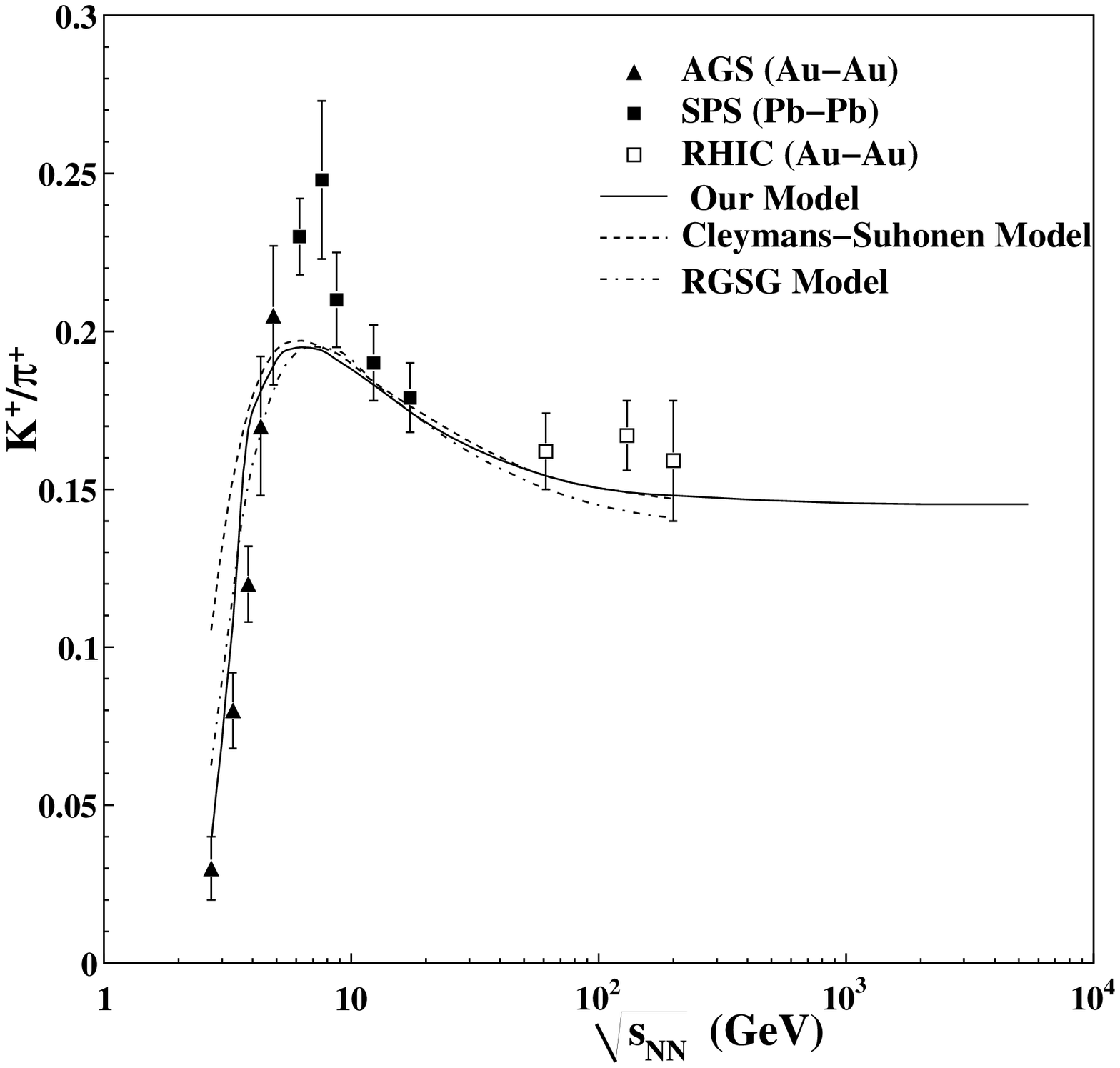}
\caption{\label{label}Variation of $K^+/\pi^+$ ratio with respect to $\sqrt{s_{NN}}$. Solid line is the results of our model \cite{Tiwari:2012} and dashed and dash-dotted lines are the results calculated in Cleymans-Suhonen model \cite{Cleymans:1987} and RGSG model \cite{Rischke:1991}, respectively. Symbols are the experimental data \cite{Tiwari:2012}.}
\end{minipage} \hspace{8pc}%
\begin{minipage}{14pc}
\includegraphics[width=14pc]{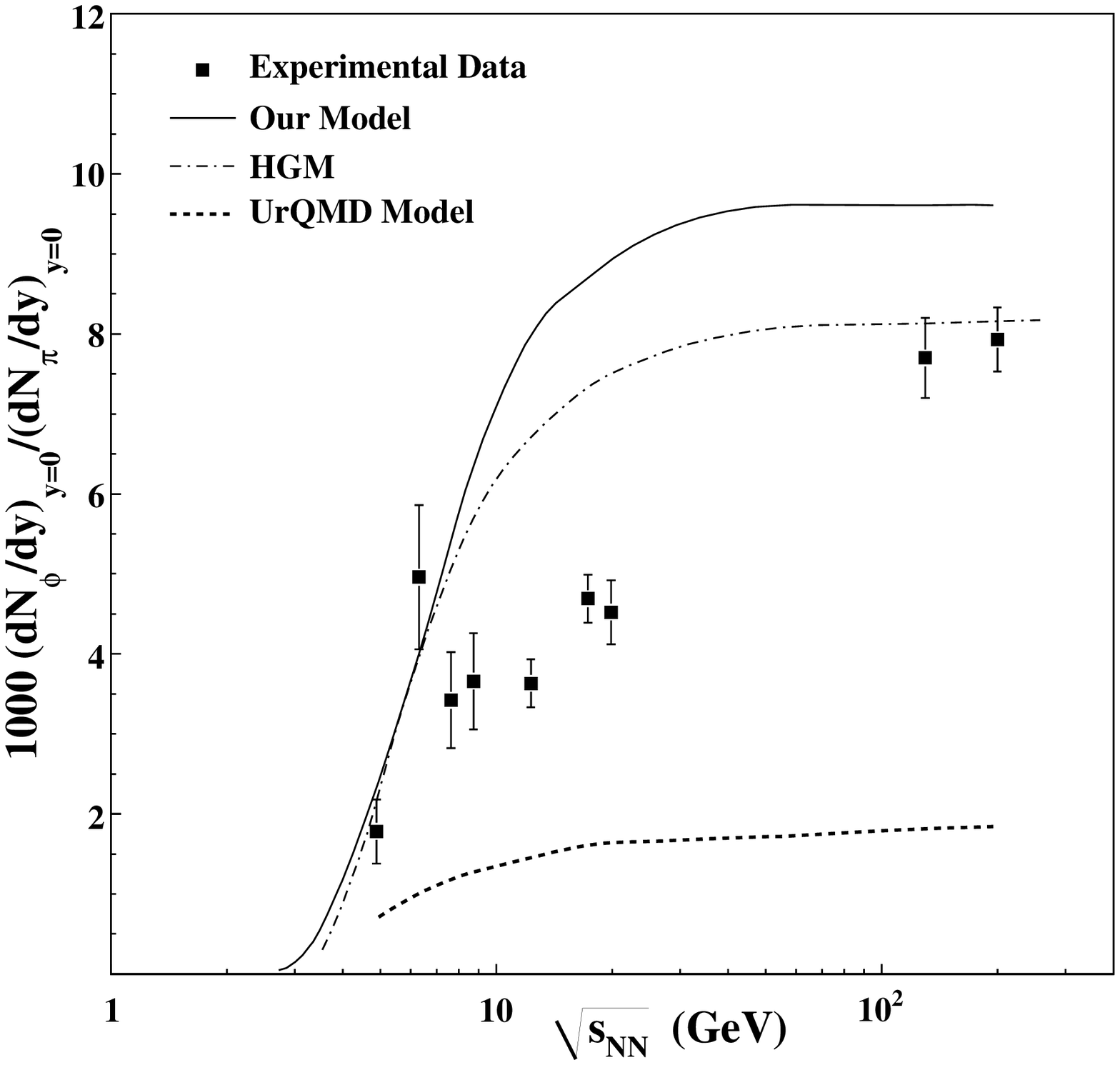}
\caption{\label{label} Energy dependence of the $\phi/\pi$, $(\pi=1.5(\pi^+ +\pi^-))$ ratio at midrapidity. Dash-dotted line is the result of statistical hadron gas model (HGM) \cite{Andronic:2006}. Solid line represents our model calculation \cite{Tiwari:2012} and dashed line is the result of UrQMD model \cite{Bleicher:1999}. Symbols are the experimental data points \cite{Blume:2011}.}
\end{minipage} 
\end{figure}

Figure 2 represents the variation of $K^+/\pi^+$ ratio with respect to center-of-mass energies ($\sqrt{s_{NN}}$). We have compared our model results with that of other models \cite{Cleymans:1987,Rischke:1991} and find that our model result is closer to experimental data in comparison to other models. We have shown the variations of the $\phi/\pi$, $(\pi=1.5(\pi^+ +\pi^-))$ ratio at midrapidity ($y=0$) with $\sqrt{s_{NN}}$ in Fig. 3. We compare our results with the experimental data \cite{Blume:2011} and statistical hadron gas model (HGM) \cite{Andronic:2006}. We find that our model provides a good description of the experimental data at lower energies but fails to describe the data at higher energies. HGM model, which has included the strangeness saturation factor $\gamma_s$ to account for the partial equilibration of strangeness, describes the data well at the lower and upper energies while it completely disagrees with the experimental data at the intermediate energies. We also show the results of the hadronic transport model Ultra-relativistic Quantum Molecular Dynamical (UrQMD) \cite{Bleicher:1999} in which kaon coalescence mechanism is used for the production of $\phi$. The predictions of UrQMD lie well below the experimental data at all energies which suggests that this mechanism fails badly in explaining the production mechanism of $\phi$.

\begin{figure}[h]
\begin{minipage}{14pc}
\includegraphics[width=14pc]{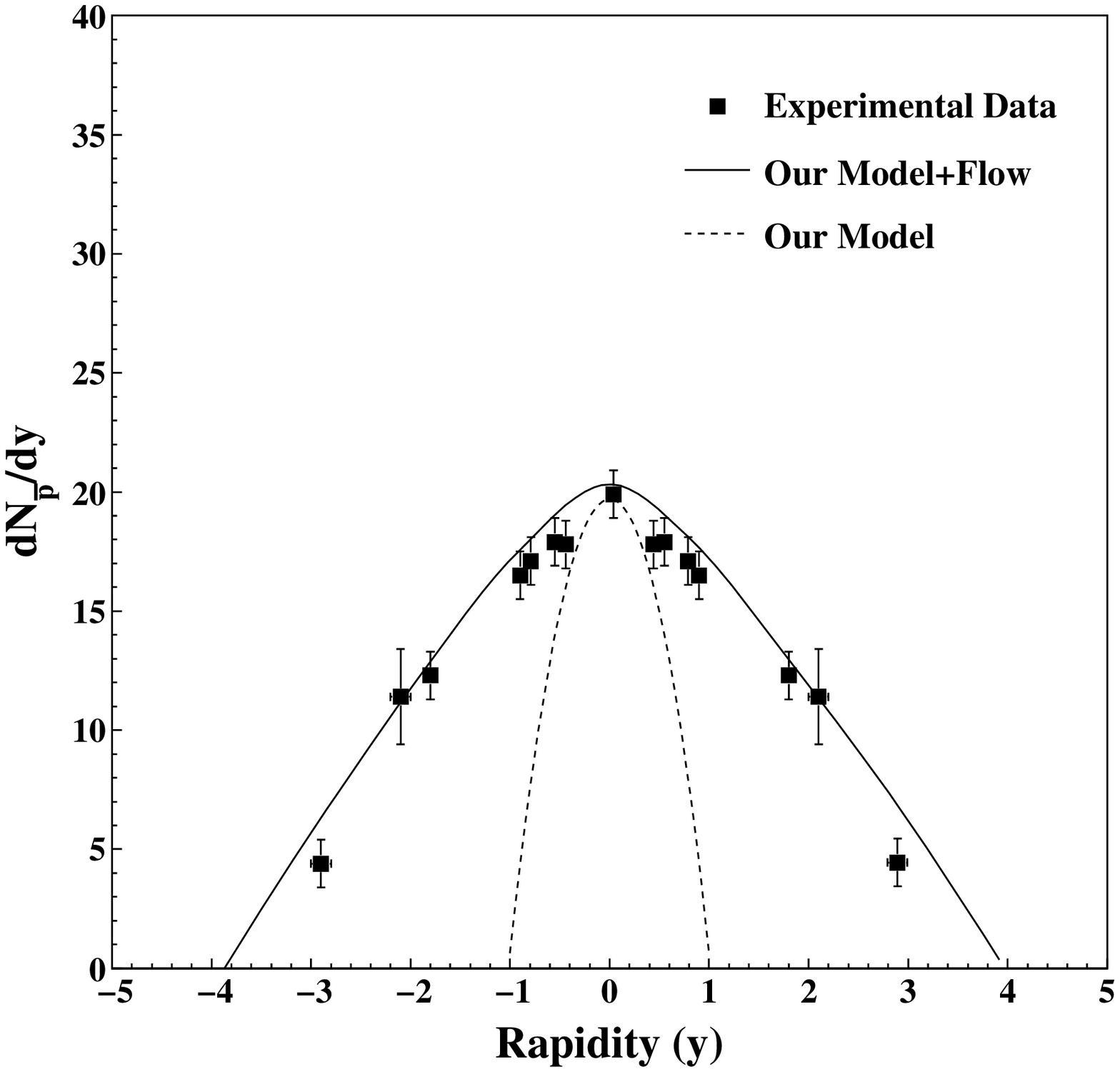}
\caption{\label{label}Rapidity distributions for $\bar{p}$ in central $Au-Au$ collisions at $\sqrt{s_{NN}}= 200\; GeV$.}
\end{minipage}\hspace{8pc}%
\begin{minipage}{14pc}
\includegraphics[width=14pc]{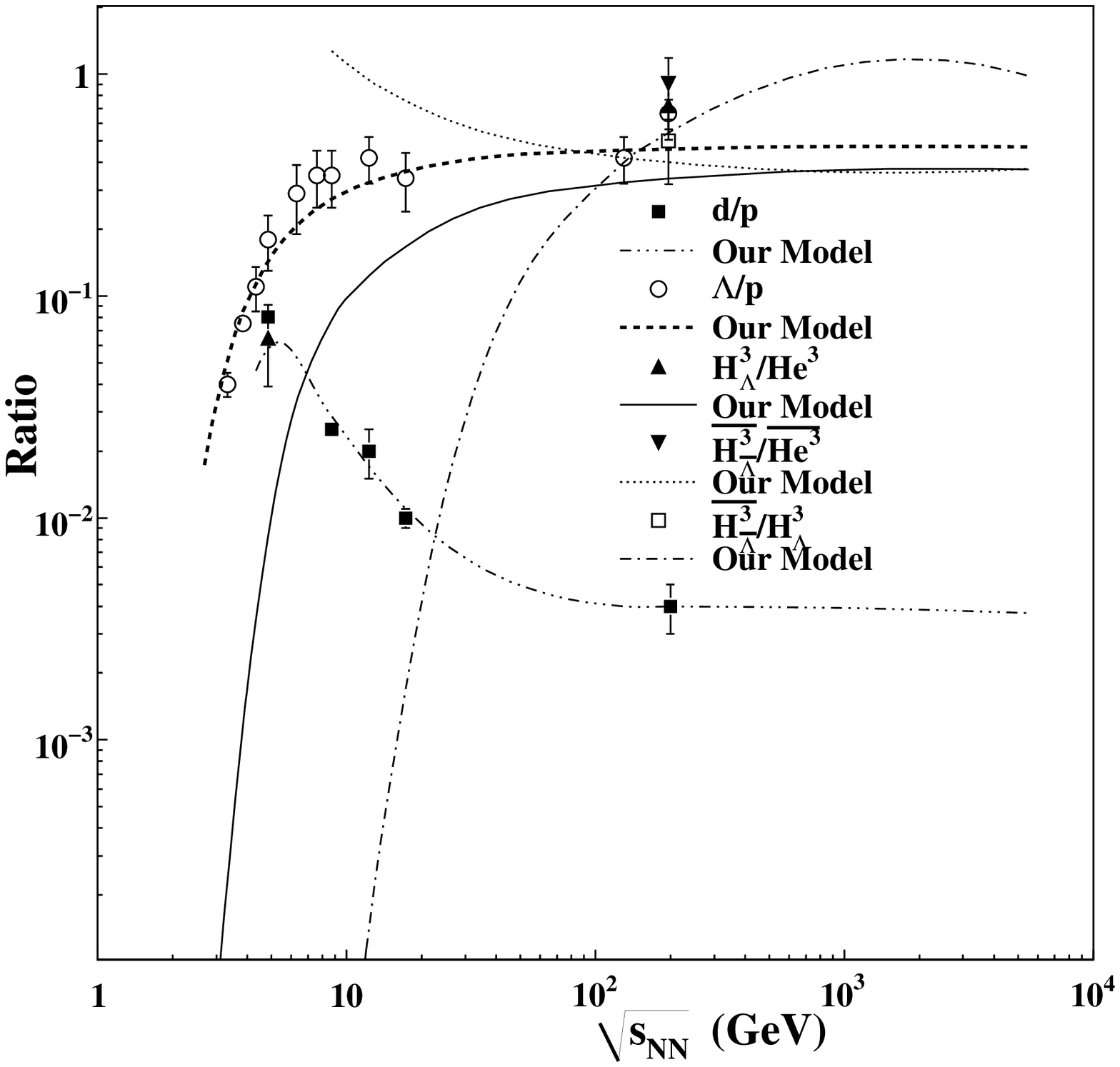}
\caption{\label{label} The energy dependence of various baryons, nuclei, and antinuclei yield ratios. Lines are our model calculations and symbols represent the experimental data \cite{Tiwari1:2013}.}
\end{minipage} 
\end{figure}

In Fig. 4, we present the rapidity distribution of anti-proton ($\bar{p}$) for central $Au-Au$ collisions at $\sqrt{s_{NN}}=200\; GeV$ over full rapidity range. Dotted line shows the distribution of $\bar{p}$ due to stationary thermal source which describes only the midrapidity data while it fails to describe the experimental data at other rapidities. Solid line shows the rapidity distributions of $\bar{p}$ after incorporation of the longitudinal flow in our thermal model and it gives a good agreement with the experimental data \cite{Bearden:2004}. After fitting the experimental data, we get the value of $\eta_{max}=3.20$ and hence the longitudinal flow velocity $\beta_L=0.922$ at $\sqrt{s_{NN}}=200\; GeV$. In Fig. 5, we show the variations of $\Lambda/p, d/p$, $H_{\Lambda}^3/He^3$ and $\bar H_{\bar\Lambda}^3/\bar He^3$ with $\sqrt{s_{NN}}$. Comparisons with the experimental data yield a good agreement between them \cite{Tiwari1:2013}.

\section{Conclusions}
In conclusion, we find that our model provides a good fit to the experimental data on ratios of hadron yields as well as mid-rapidity densities of various particles. We present an analysis of rapidity distributions of hadrons at various center-of-mass energies using our model. We see that our thermal model alone cannot describe the experimental data fully unless we incorporate flow velocity in the longitudinal direction and as a result our modified model predictions show a good agreement with the experimental data. Although, thermal model fails in case of multistrange particles which suggests that a somewhat different type of mechanism e.g., coalescence model is required. Further, we have analyzed the production of nuclei and hypernuclei in terms of our model which describe the experimental data very well.

\section{Acknowledgments}
SKT is grateful to Council of Scientific and Industrial Research (CSIR), New Delhi for providing a research grant.

\smallskip

\end{document}